\documentstyle[12pt,twoside,fleqn,espcrc1,epsf]{article}

\newcommand{\AmS}{{\protect\the\textfont2
  A\kern-.1667em\lower.5ex\hbox{M}\kern-.125emS}}
     
\title{Au+Au Reactions at the AGS:
Experiments E866 and E917}

\author{C.A. Ogilvie for the E866 and E917 Collaborations
\address{Department of Physics, MIT, Cambridge, MA 02139, USA}}
\begin{document}
\input epsf
\epsfverbosetrue
% typeset front matter
\maketitle

E866 Collaboration\\
L.Ahle$^9$, Y.Akiba$^6$, K.Ashktorab$^2$, M.D.Baker$^9$, D.Beavis$^2$,
H.C.Britt$^8$, J.Chang$^4$,
 C.Chasman$^2$, Z.Chen$^2$, 
Y.Y.Chu$^2$
% ...
\\
$^1$ Argonne National Laboratory, %Argonne, IL 60439\\
$^2$ Brookhaven National Laboratory, %Upton, NY 11973 \\
$^3$ University of California, Space Sciences Laboratory, Berkeley, %CA 94720 \\
$^4$ University of California, Riverside, %CA 92507 \\
$^5$ High Energy Accel. Res. Organization (KEK), %Oho, %Tsukuba, Ibaraki 305,
%Japan\\
$^6$ High Energy Accel. Res. Organization (KEK), Tanashi-branch, %Midoricho,
%Tanashi, Tokyo 188, Japan\\
$^7$ University of Illinois at Chicago, %Chicago, 60607 \\
$^8$ Lawrence Livermore National Laboratory, %Livermore, CA 94550 \\
$^9$ Massachusetts Institute of Technology, %Cambridge, MA 02139 \\
$^{10}$ Department of Physics, University of Tokyo, %Tokyo 113, Japan \\
$^{11}$ University of Tsukuba, %Tsukuba, Ibaraki 305, Japan \\
$^{12}$ Yonsei University, %Seoul 120-749, Korea\\
$^{13}$ University of Maryland, %College Park, MD 20742\\
$^{14}$ University of Rochester, %Rochester, NY 14627 \\
$^{15}$ Center for Nuclear Study, School of Science, University of Tokyo, 
%Midoricho, Tanashi, Tokyo 188, Japan\\
$^{16}$ Oak Ridge National Laboratory, %Oak Ridge, Tennessee 37831\\
$^{17}$ Columbia University %New York 10027 and Nevis Laboratories, 
%Irvington, New York  10533\\

\begin{abstract}
Particle production and
correlation functions from Au+Au reactions have been measured as a function of
both beam energy (2-10.7AGeV) and impact parameter.
These results are used to probe the dynamics of heavy-ion reactions,
confront hadronic models over a wide range of conditions and
to search for the onset of new phenomena.  
\end{abstract}

\section{INTRODUCTION}

Very dense nuclear matter is formed in Au+Au collisions at beam energies 
near 10 AGeV.  The physics of such dense matter is intriguing - 
whether the system behaves as 
an interacting mixture of excited hadronic resonances, or whether   
the 
density is sufficiently high that some of the matter is converted to 
deconfined quarks and gluons.

Experiments E866 and E917 at the AGS aim to characterize as fully as 
possible the emission of particles from 
Au+Au collisions to probe the properties of such dense matter.  In Quark Matter 
'96\cite{qm96} the E866 collaboration 
reported that the proton rapidity distribution for central 
collisions is peaked at mid-rapidity\cite{Chen98}.  This is
consistent with forming a dense, baryon-rich system.
The dynamics of this system were further 
characterized by the spectra and yield of the most abundantly 
produced particles, pions and 
kaons.   These results are 
summarized here for reference. The measured mid-rapidity K$^+/\pi^+$ 
ratio in 
central reactions is 0.19$\pm$0.01.  This continues a steady increase in 
the K$^+/\pi^+$ ratio from p+A\cite{pA} and Si+A\cite{SiA} 
reactions. The transverse $m_t$ spectra of  
different particles have different spectral shapes, with an overall trend 
that 
the proton spectra have larger inverse slopes than the kaon spectra, which in
turn have larger inverse slopes 
than the pion spectra.  Detailed analysis\cite{qm96,Chen98} 
revealed that the $\pi^-$ transverse spectra 
have a low p$_t$ rise above an exponential that is steeper than the rise
in $\pi^+$ spectra.  
Such features 
originate from the interplay between multiple collisions
of hadrons, expansion of the system, decay of resonances 
and Coulomb effects.

Since Quark Matter '96, the E866 and E917 collaborations have 
continued to work on
several open questions, two of which will be 
the main topics of this paper; 1) is there any evidence for new physics beyond 
hadronic multiple collisions, for example the formation of a baryon rich 
quark-gluon plasma? and 2) is the collision zone at thermal 
equilibrium?  These questions are linked because a high 
collision rate will drive the system towards equilibrium and
also alter the signatures of a possible quark-gluon plasma 
that might be formed early in the reaction.

\section{EXPERIMENT}

The results in this talk come from two separate collaborations, E866 and 
E917, that used much of the
same equipment.  Experiment E866 measured
 Au+Au reactions at 2, 4, and 10.7 A GeV 
kinetic 
energy.  
Experiment E917 also measured Au+Au reactions but at 6, 8 and 10.7 
A GeV kinetic energy.  
Experiment E866 consists of two magnetic spectrometers that rotate
independently.
The Forward spectrometer was used only in E866 and 
was optimized for tracking in the high-multiplicity forward region. 
This 6msr spectrometer  
consists of a sweeping magnet
followed by two tracking stations before and after a dipole magnet.  
Each tracking station is a TPC placed between high resolution drift 
chambers. Particle identification is performed with a TOF-wall 
with 70ps timing resolution.
The older spectrometer, Henry Higgins, has an
acceptance of 25msr and was used by both E866 and E917.  
It consists of drift chambers on either side of a dipole magnet. 
Particle identification is performed with
a TOF-wall with a timing resolution of 130ps.  
The Henry Higgins spectrometer has an on-line particle identification 
trigger.
Both experiments used the same detectors for global characterization of the 
events:  
a multiplicity counter surrounding the target, and a calorimeter placed 
at zero degrees with a 1.5$^o$ opening angle. E917 added a beam-vertexing 
detector to improve the reaction plane measurement.   
The total systematic uncertainty in the 
normalization of the cross-sections is 15$\%$ 
and
the systematic uncertainty on the inverse slope parameters
is estimated to be 5$\%$.
For more details on the equipment and data analysis the
reader is referred to references\cite{Chen98,nim}.

\section{EXCITATION FUNCTION}

Measurement of an excitation function can provide more
insight into the mechanism of particle production
than results at a single beam energy.
An excitation function can also confront hadronic cascade models
over a wider range of conditions 
and thus permit a search for the onset of any new phenomena,
e.g. the formation of a small volume of a baryon-rich QGP.

\begin{figure}[htb]
\begin{minipage}[t]{75mm}
\epsfxsize=7.5cm\epsfbox[90 300 440 585]{
%earth$scratch:[ogilvie.qm97]mt_excit.eps}
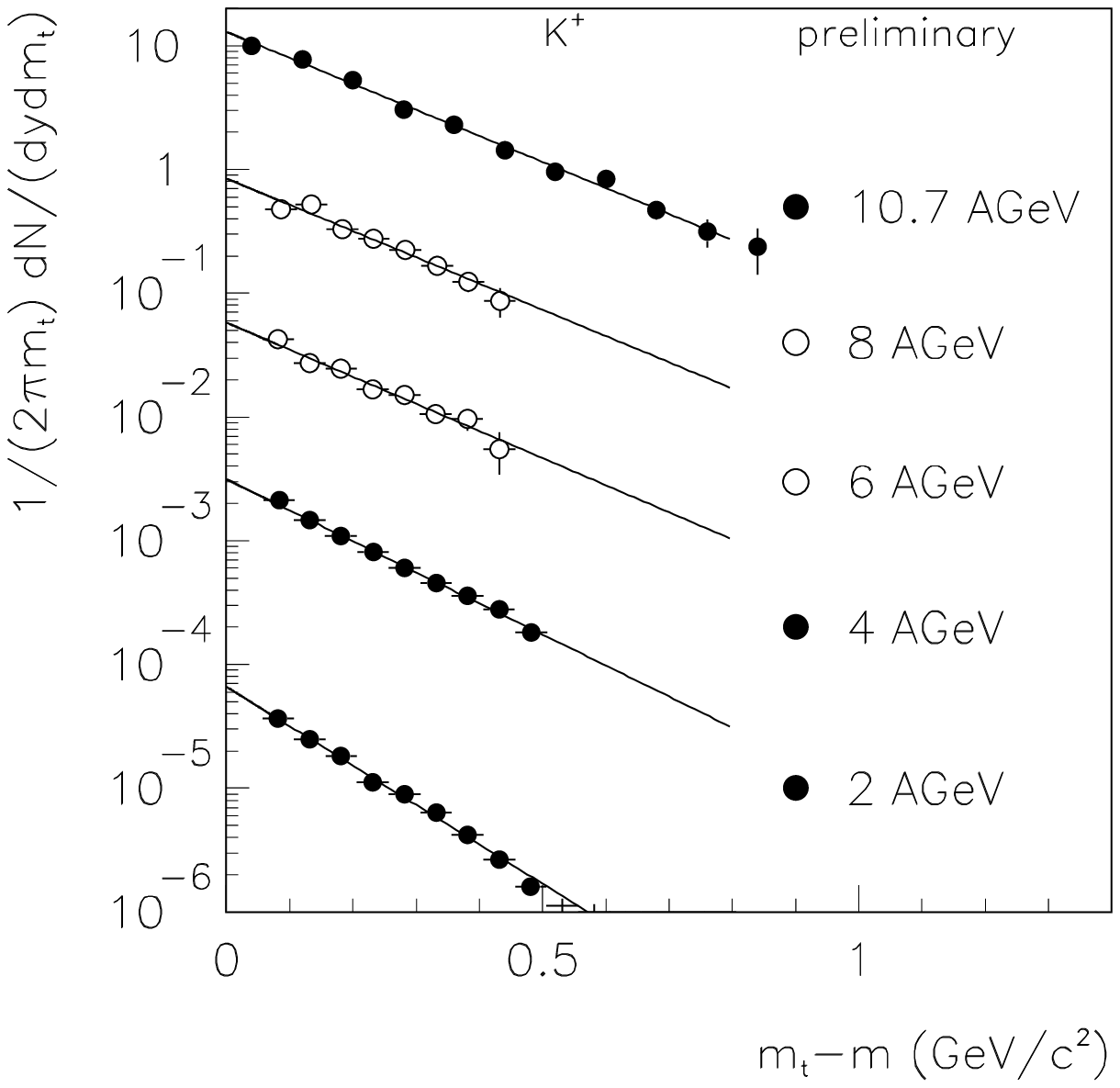}
\end{minipage}
\hspace{\fill}
\begin{minipage}[t]{75mm}
\epsfxsize=7.5cm\epsfbox[90 300 440 585]{
%earth$scratch:[ogilvie.qm97]mt_pion.eps}
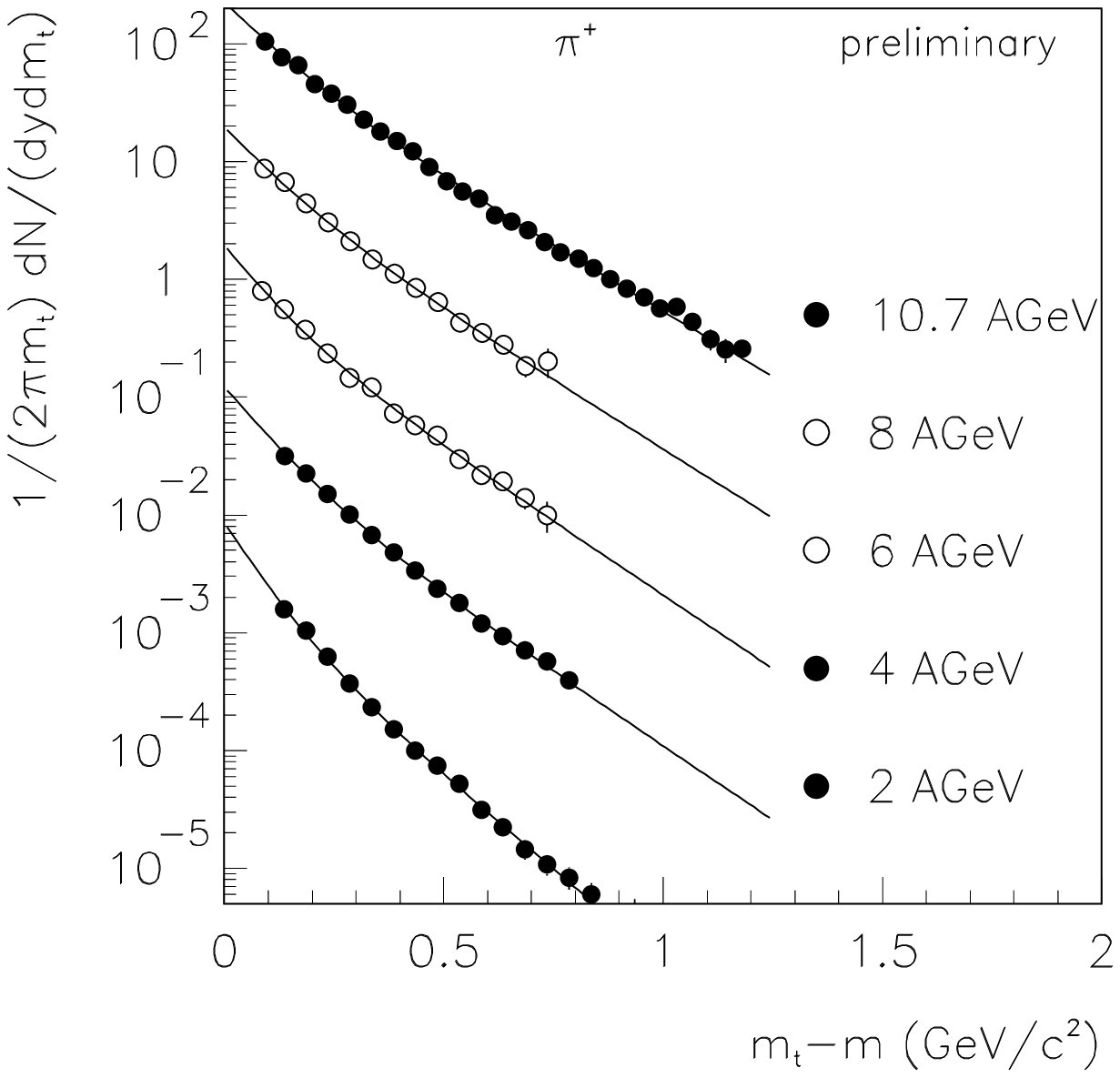}
\end{minipage}
\caption{The invariant yield of K$^+$ (left panel) and $\pi^+$
 (right panel) per event as a function 
of m$_t$ at mid-rapidity for the different beam energies.
The data from E866 are shown as filled circles and the data from
E917 are shown as open circles. The data at 10.7AGev are shown at the
correct scale, the data at each lower energy are 
divided by ten for clarity.
The errors are statistical only.
}
\label{fig:mtkaon}
\end{figure}  

Kaon production is a sensitive probe of hadronic
multiple 
collisions in heavy-ion reactions and may also provide 
evidence for the formation of a 
QGP.  The left panel of Figure \ref{fig:mtkaon} shows a 
plot of the invariant yield of kaons 
as a function of transverse 
mass, m$_t=\sqrt{p_t^2+m_0^2}$, produced in
Au+Au central collisions at 2, 4, 6, 8, and 10.7 AGeV kinetic beam 
energy.
The centrality selection at each beam energy was the inner 8\%
of the total interaction cross-section ($\sigma_{int}$=6.8b).
Each spectrum covers 
the mid-rapidity range corresponding
to that beam energy, $|\frac{y-y_{nn}}{y_{nn}}|<0.25$.  
The kaon spectra were fit
with an exponential function in m$_{t}$.
%\begin{equation}
%\frac{1}{2\pi m_t}\frac{d^2N}{dm_tdy}=
%\frac{dN/dy}{2\pi(Tm_0 + T^2)}e^{-(m_t-m_0)/T} 
%\end{equation}
The fits reproduce the spectra well and
provide the inverse slope
parameter T and the rapidity density, dN/dy, in that
rapidity slice.
 
The right panel of Figure~\ref{fig:mtkaon} 
shows the invariant spectra for $\pi^+$ from 
Au+Au reactions at each of the five beam energies.
The pion spectra  
rise at low p$_t$ above an exponential.  These
spectra have therefore been fitted with a double exponential,            
with the dN/dy and mean m$_t$ as two of the fit parameters. 
%The other two
%parameters are the relative normalizations of the exponentials and the
%difference of the two inverse slope parameters.                           

\begin{figure}[htb]
\epsfysize=8cm\epsfbox[1 130 565 505]{
%earth$scratch:[ogilvie.qm97]yield_mt_ebeam.eps}
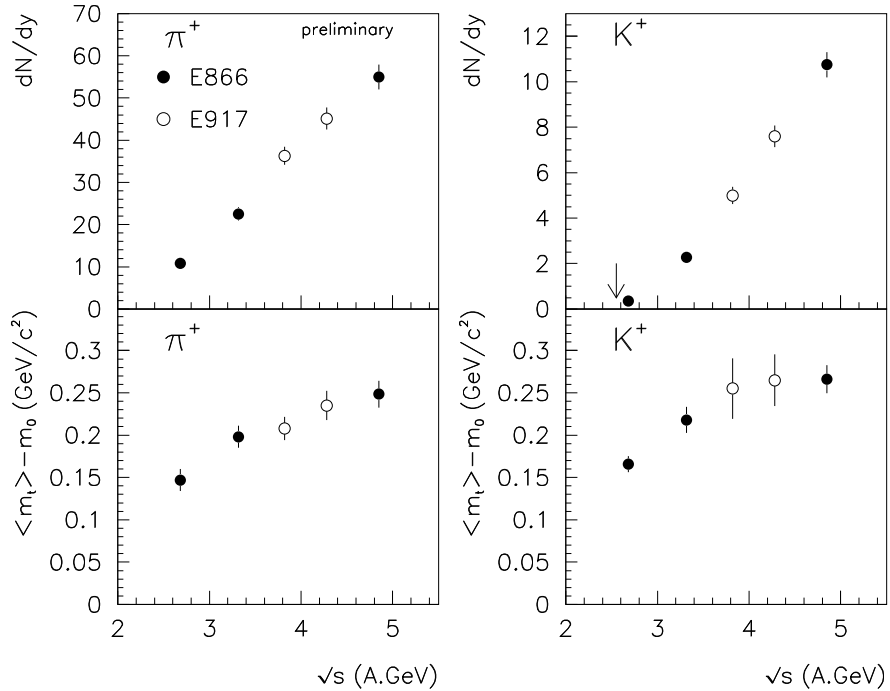}
\caption{The yield of $\pi^+$ and K$^+$ at mid-rapidity (top-panels)
for central Au+Au
reactions as a function of the initial available beam energy. The lower panels
show the mean m$_t$ minus the rest mass
for $\pi^+$ and K$^+$ at 
the same rapidity.
The arrow indicates the N-N threshold for K$^+$ production.
The errors include both statistical and a 5\% point-to-point
systematic uncertainty.
}
\label{fig:excit}
\end{figure}

The mid-rapidity yields of $\pi^+$ and 
K$^+$ as a function of the initial available energy $\sqrt{s}$ are 
shown in the upper panels
of Figure~\ref{fig:excit}.  
Both  
pion and kaon yields increase steadily and smoothly with beam energy.   
There is no indication of any sudden increase in particle yield with increasing
$\sqrt{s}$.  
In the lower panels of Figure~\ref{fig:excit}, 
the mean m$_t$ of the transverse 
spectra for pions and kaons are plotted versus $\sqrt{s}$.  
Compared to the increase 
of particle production, the mean m$_t$ increases
slowly with beam 
energy. The extra available energy mainly goes into more particle
production rather than increased transverse energy.                

\begin{figure}[htb]
\begin{minipage}[t]{75mm}
\epsfxsize=7.5cm\epsfbox[30 160 520 495]{
%earth$scratch:[ogilvie.qm97]kpi_ebeam_model.eps}
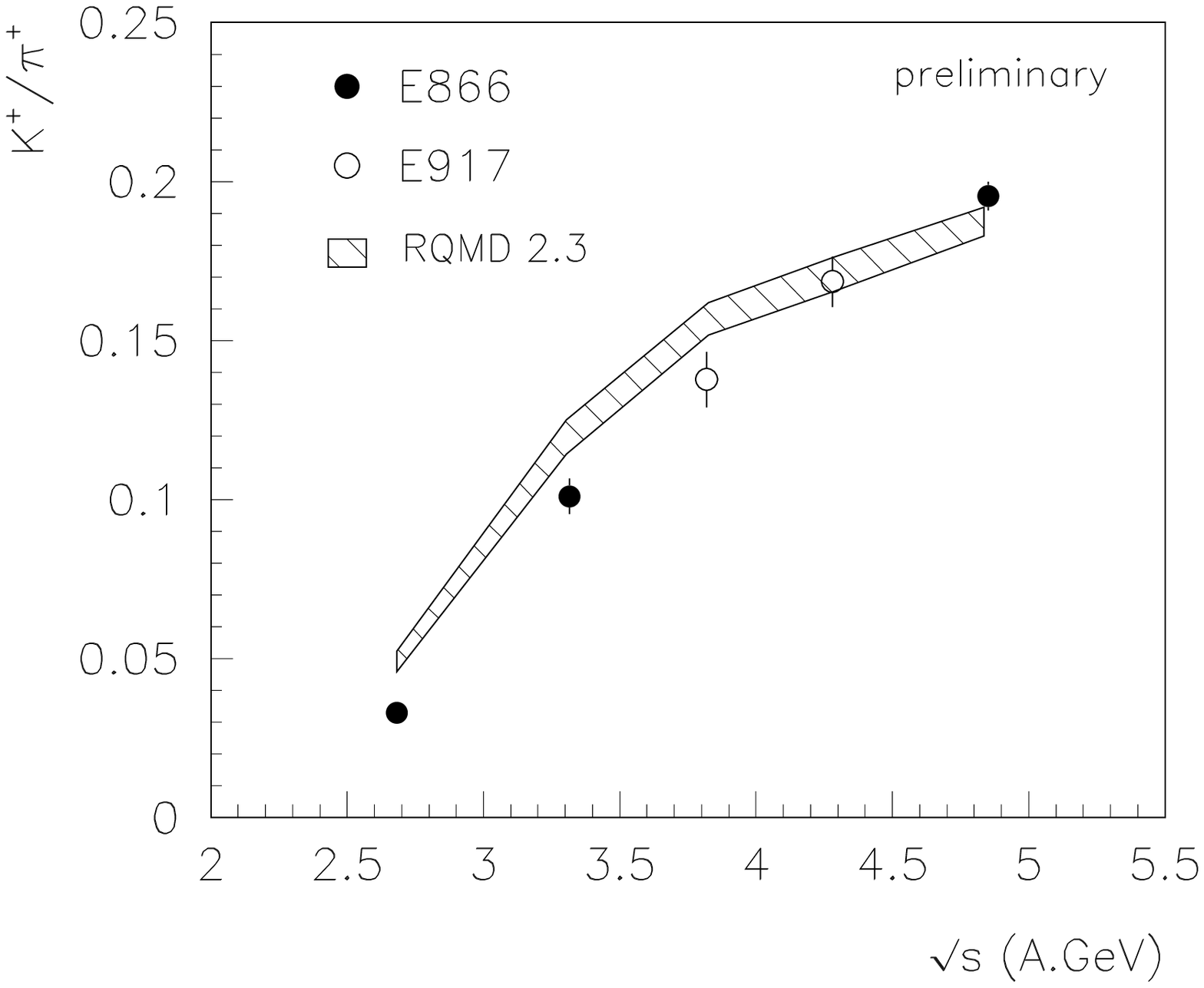}
\caption{The ratio K$^+$/$\pi^+$ at mid-rapidity in
central Au+Au reactions as a function 
of the initial available energy.
The errors are statistical only.
%The data from E866 are shown as solid circles and the data from
%E917 are shown as open circles.}
}
\label{fig:kpi}
\end{minipage}
\hspace{\fill}
\begin{minipage}[t]{75mm}
\epsfxsize=7.5cm\epsfbox[30 160 520 495]{
%earth$scratch:[ogilvie.qm97]enhance.eps}
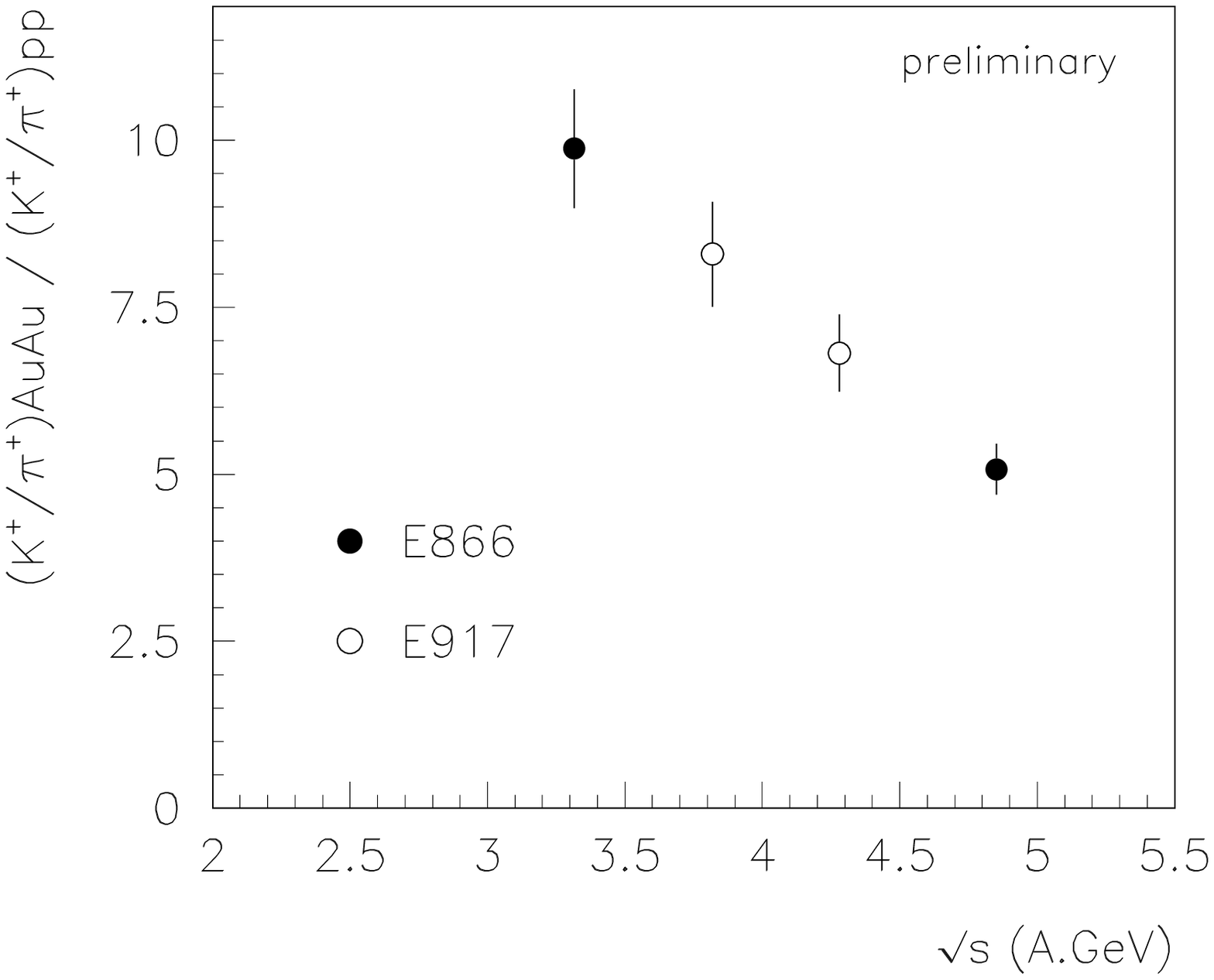}
\caption{The double ratio K$^+$/$\pi^+$ at mid-rapidity from central
Au+Au reactions divided by K$^+$/$\pi^+$ of total yields from p+p
reactions as a function 
of the initial available energy.
The errors include both statistics and a 5\%
systematic uncertainty in the K/$\pi$ ratio from p+p reactions.
%The data from E866 are shown as solid circles and the data from
%E917 are shown as open circles.}
}
\label{fig:enhance}
\end{minipage}
\end{figure}

The increase in kaon yield 
with beam energy is more rapid than the increase of pion yield.  
This is emphasized in 
Figure~\ref{fig:kpi}, 
where the $K^+/\pi^+$ ratio is plotted versus $\sqrt{s}$.  
This ratio increases steadily from 
near 3\% at 2 AGeV to 19\% at 
10.7 AGeV. 

These extensive measurements over a broad range of conditions
provide a stringent test of 
hadronic models.  The predictions of the 
RQMD hadronic 
cascade\cite{Sorge89} have been compared to the 
measured ratio K$^+/\pi^+$ (Figure~\ref{fig:kpi}).  
The model reproduces the data well.
  
It is however hard to distill any simple
physics lesson from a comparison with such a complicated model.
One way to improve our understanding is to divide the Au+Au K$^+/\pi^+$ ratio
by the K$^+/\pi^+$ ratio from p+p 
reactions\cite{Ross75,Fes79,Blob} (Figure~\ref{fig:enhance}).
This double ratio is greater than one, demonstrating that
K/$\pi$ is enhanced in Au+Au 
reactions relative to p+p collisions.  
This enhancement is largest at the lowest beam 
energy, possibly because secondary collisions 
increase in relative 
importance
compared to initial collisions as the beam energy is reduced.
At the kaon threshold this double ratio goes to infinity\cite{caveats}.
Also note that the double ratio at
2GeV is not plotted because of the large uncertainty
in the p+p kaon yield.
   
\section{CENTRALITY DEPENDENCE}

At each beam energy the experiment was triggered primarily 
with spectrometer-based 
triggers.  
The spectrometer data can then be sorted
off-line into several centrality classes.  
At this stage of the analysis, results are
only available for the centrality dependence of particle
production at the full energy, 11.6 A~GeV/c, from Experiment E866.

As  the  reactions become more central, 
the number of primary and secondary hadron-hadron
collisions increase. 
By measuring particle spectra and yields as a function of centrality
we probe how 
these collisions affect particle production.
In addition, because the volume of dense nuclear matter increases
with centrality\cite{ART}, we can 
search for the onset of any new phenomena associated with high densities.

The measured energy near zero
degrees (E$_{ZCAL}$) is  dominated by 
projectile spectator nucleons and  provides an estimate
of the initial overlap of the projectile and target nuclei.
The number of projectile
participants, N$_{pp}$, is estimated from the measured 
E$_{ZCAL}$  
\begin{equation}
\rm{N}_{pp}=197\times (1 - \rm{E}_{ZCAL}/E^{kin}_{beam})
\end{equation}
where $E^{kin}_{beam}$=2123 GeV is the kinetic energy of the beam.
The event classes are defined by cuts in E$_{ZCAL}$.
These are referenced by their cross-section:
0-5\%, 5-12\%, 12-23\%, 23-39\% and 39-73\% of the total interaction
cross-section, $\sigma_{int}$=6.8b.

\begin{figure}[htb]
\epsfxsize=12cm\epsfbox[3 120 566 500]{
%earth$scratch:[ogilvie.qm97]kaon_prod.eps}
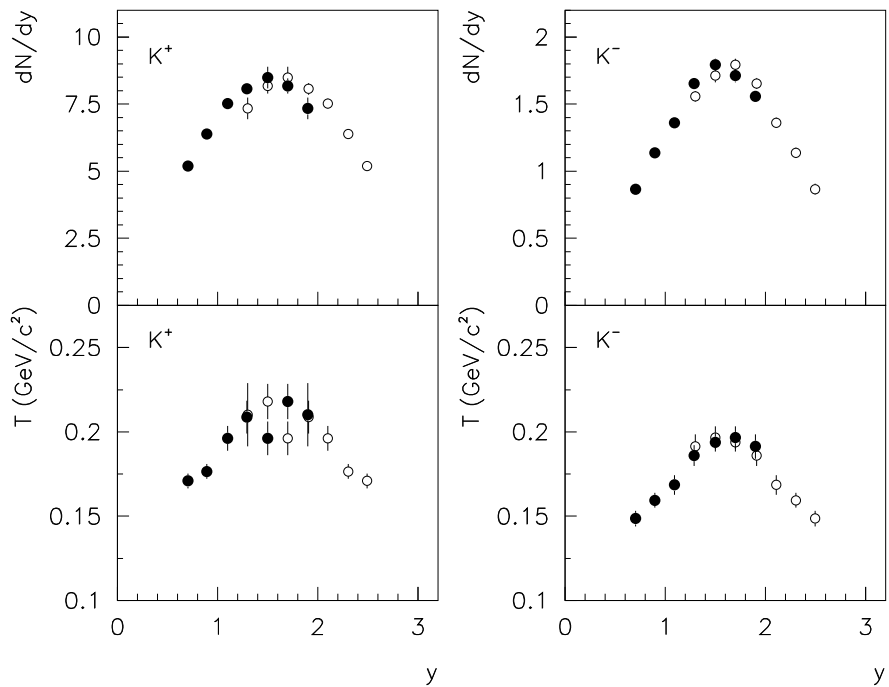}
\caption{The dN/dy and inverse slope
distributions as a function of rapidity for
K$^+$ and K$^-$ emitted from mid-central
Au+Au reactions at 11.6 A~GeV/c.
The errors are statistical only.
The hollow circles are the data points reflected
about mid-rapidity (y=1.6).}
\label{fig:kdndy}
\end{figure}  
In the top panels of Figure~\ref{fig:kdndy}, dN/dy distributions 
for K$^+$ and K$^-$ are shown for the 5-12\% event class
of Au+Au reactions at 11.6 AGeV/c. 
These have been extracted from fits to invariant spectra as a function
of transverse mass, similar to those shown in
Figure~\ref{fig:mtkaon}.
Also extracted from each transverse spectrum is
the inverse slope,  T. The rapidity dependence
of the inverse slope parameter
is shown in the lower panels of
Figure~\ref{fig:kdndy}.
Both the dN/dy and the inverse slope distributions peak at 
mid-rapidity.  The dN/dy distribution for K$^-$ 
is narrower than for K$^+$, 
and the K$^+$ inverse 
slopes tend to be slightly larger than 
the inverse slope for K$^-$.  

\begin{figure}[htb]
\begin{minipage}[t]{100mm}
\epsfxsize=10.cm\epsfbox[30 120 460 395]{
%earth$scratch:[ogilvie.qm97]kpernpp.eps}
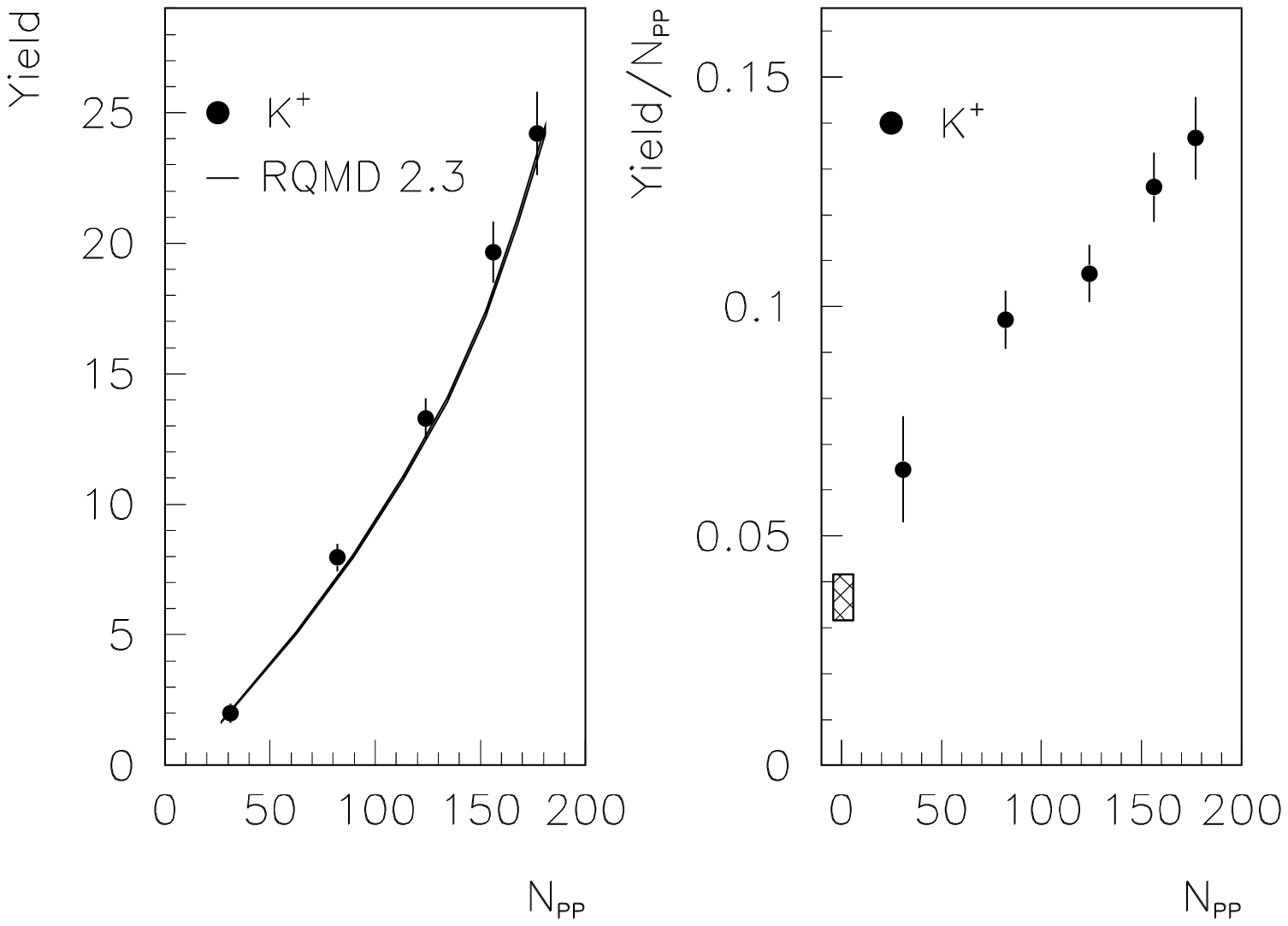}
\end{minipage}
\hspace{\fill}
\begin{minipage}[t]{47.5mm}
\epsfxsize=4.75cm\epsfbox[35 120 240 395]{
%earth$scratch:[ogilvie.qm97]ratnpp.eps}
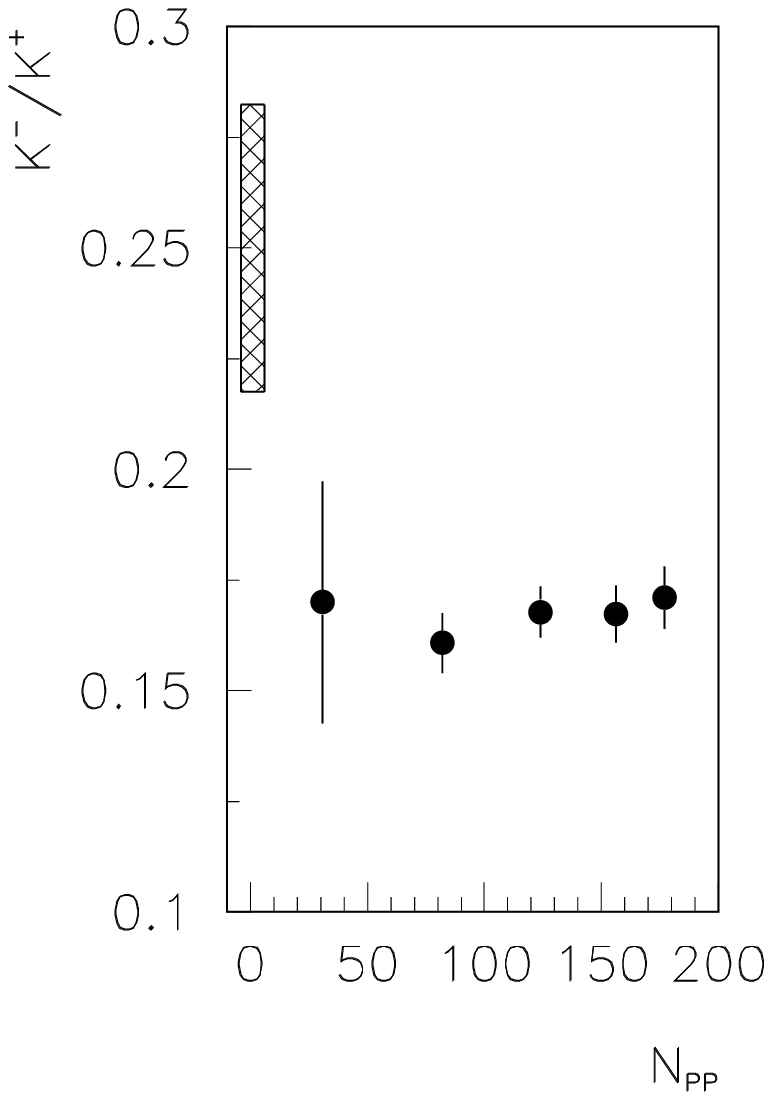}
\end{minipage}
\caption{Left panel: the average yield of kaons in an event as a function
of the number or projectile participants
in Au+Au reactions at 11.6 A~GeV/c.
Middle panel: the yield of K$^+$ per projectile 
participant. 
Right panel:
The ratio of the 
total yields of K$^-$ and K$^+$ as a function of the number of projectile 
participants.
The errors are statistical only. The hatched boxes are the
yields from the isospin averaged initial N-N collisions.} 
\label{fig:kyield}
\end{figure}

To extract the total yield
of kaons per event, 
each dN/dy distribution was fitted with a Gaussian centered at 
mid-rapidity (y=1.6) 
In Figure~\ref{fig:kyield}, the K$^+$ yield is plotted  
as a function of the number of projectile participants.  
The yield of K$^+$ increases non-linearly with
N$_{pp}$, 
implying that kaon production is more efficient in central collisions.  
This is emphasized in the middle panel of Figure~\ref{fig:kyield}, 
where the K$^+$ yield 
per participant is plotted.  
The yield of kaons per participant in central 
reactions is 3.5 times the yield
from isospin-averaged N-N collisions at the same beam energy\cite{Fes79,Blob}.
This suggests that the majority of kaons in 
central reactions are produced by secondary collisions.

The hadronic cascade model, RQMD\cite{Sorge89}, attempts
to simulate how these 
secondary collisions occur in a heavy-ion reaction.
The model reproduces both the absolute magnitude of 
kaon production and the dependence of yield on the number of projectile 
participants quite well.

The yields of $\pi^+$, $\pi^-$, K$^-$,
and $\overline{\rm{p}}$ have also been measured as a function of centrality.
The right panel of Figure~\ref{fig:kyield} 
shows the ratio of total yields K$^-$ to K$^+$. 
This ratio is strikingly constant over the measured range of 
centrality and below the estimated K$^-$/K$^+$ ratio
from initial N-N collisions\cite{Fes79,Blob}.  
The K$^-$ yield increases non-linearly with N$_{pp}$ 
in a manner very similar to 
the K$^+$ yield.  This is surprising since the two particles have different 
underlying production mechanisms, energy thresholds and absorption 
rates.

The observation that the 
K$^-$/K$^+$ ratio is independent of centrality may also 
test models that include an in-medium mass 
for kaons.  There have been several predictions that the K$^-$ mass 
decreases with increasing 
nuclear density \cite{Weise96,Li96,Ehe96}.  
Such a conjecture could lead to more phase space for 
K$^-$+K$^+$ pair production which would increase the 
K$^-$/K$^+$ ratio in 
central reactions.  This is not observed, potentially placing a constraint
on the in-medium properties of kaons. 

\begin{figure}[htb]
\begin{minipage}[t]{75mm}
\epsfxsize=7.5cm\epsfbox[30 120 415 410]{
%earth$scratch:[ogilvie.qm97]pbarvsnpp.eps}
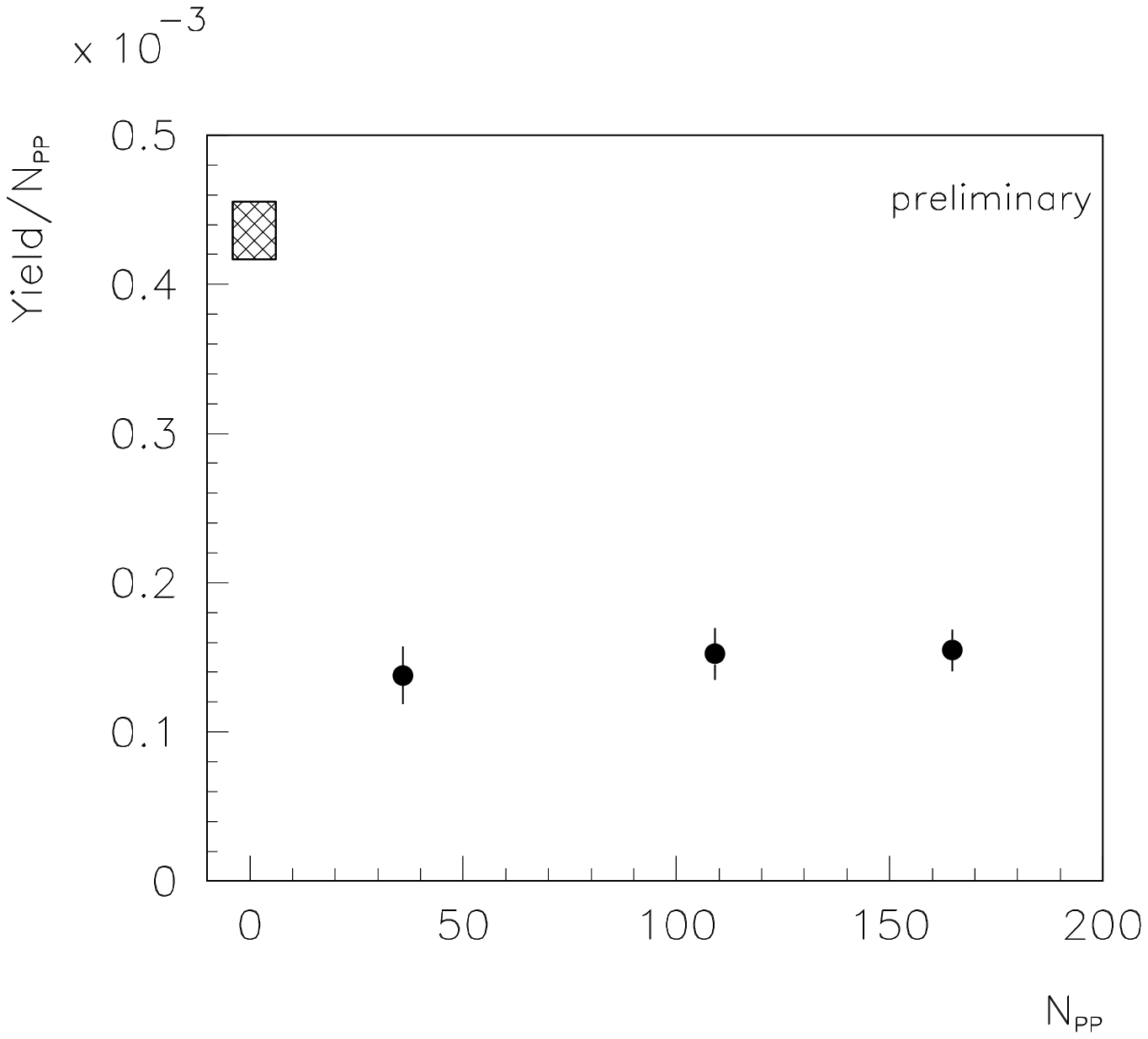}
\caption{The yield of $\overline{\rm{p}}$ per participant 
as a function of the number of projectile 
participants in Au+Au reactions at 11.6 A~GeV/c.
The errors are statistical only.
The box on the left is the estimated yield from the
initial N-N collisions.}
\label{fig:pbar}
\end{minipage}
\hspace{\fill}
\begin{minipage}[t]{75mm}
\epsfxsize=7.5cm\epsfbox[30 145 510 500]{
%earth$scratch:[ogilvie.qm97]pbar_mt.eps}
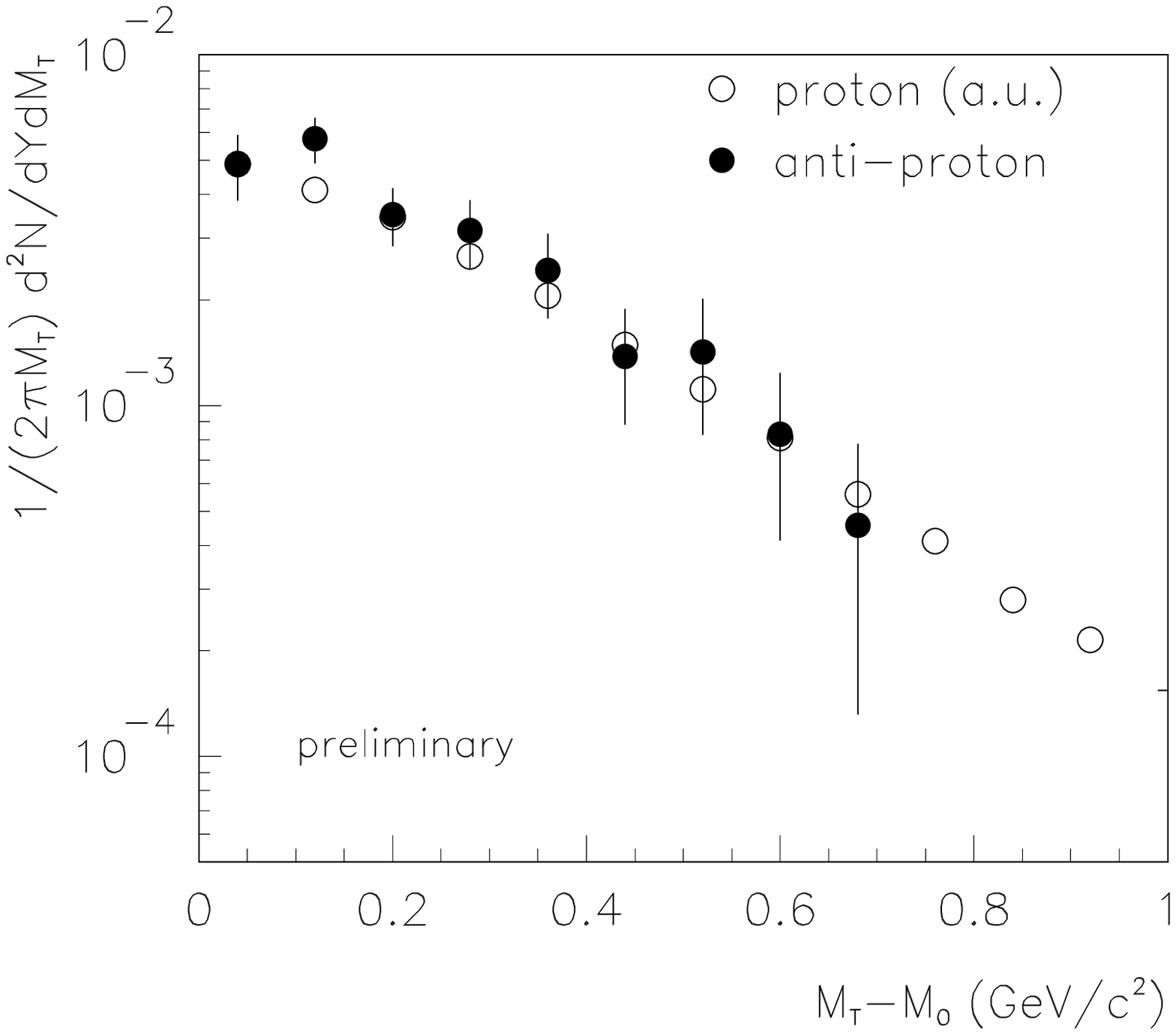}
\caption{The invariant spectrum of $\overline{\rm{p}}$ at mid-rapidity 
as a
function of transverse mass
in Au+Au reactions at 11.6 A~GeV/c.
For comparison
the proton spectrum at the same rapidity is also 
shown.
The errors are statistical only.}
\label{fig:pbarmt}
\end{minipage}
\end{figure}

The role of absorption in heavy-ion reaction and how it affects the final 
yield of particles is perhaps 
best studied with anti-protons\cite{Jahns92}. Figure~\ref{fig:pbar} is a 
plot of the total yield of $\overline{\rm{p}}$ 
per participant as a function of the 
number of projectile participants.       
The total  yield is from a Gaussian fit to
the $\overline{\rm{p}}$ rapidity distribution
measured with the Henry Higgins spectrometer\cite{feed}.

In contrast to kaon production, the $\overline{\rm{p}}$ yield 
per participant is constant and considerably below the 
N-N value\cite{Ross75}.  
Both observations are 
consistent with $\overline{\rm{p}}$ being strongly
absorbed.

The final yield of $\overline{\rm{p}}$
is sensitive not only to absorption but also to the extent
that  secondary collisions enhance $\overline{\rm{p}}$ 
production. 
Extra 
information to help disentangle these two effects
may come from the transverse spectra of 
$\overline{\rm{p}}$.  
The mid-rapidity spectrum of 
$\overline{\rm{p}}$ as measured with the Forward-spectrometer\cite{feed}
 in central 
Au+Au reactions is shown in Figure~\ref{fig:pbarmt}.   
For comparison the proton spectrum 
is shown on the same plot arbitrarily
normalized to the $\overline{\rm{p}}$ yield at low m$_t$.
Within the current statistics these two spectra have a similar shape.
This is a puzzle since $\overline{\rm{p}}$
produced at low m$_t$
should be strongly absorbed, which would
effectively increase the inverse slope parameter.
This effect may, however, be counterbalanced
by the low phase
space available for $\overline{\rm{p}}$ production which effectively
reduces the
inverse slope parameter.  

\section{PION INTERFEROMETRY}

The interference of identical particles (HBT) 
provides
additional information on the dynamics of heavy-ion reactions
and on the space-time extent of particle emission.
Experiment E866 has two parallel efforts in this field;
1) to map out the
systematic evolution of the HBT source parameters with
changing initial system size, from peripheral
Si+A reactions to central Au+Au collisions, and
2) to probe the reaction dynamics
by extracting HBT parameters as a function of 
the momentum of the particles\cite{Heinz96,Beker95}.
Only
the first aspect of our efforts will be reported here.

Experiments E859 (Si+A) and E866 (Au+Au) have measured the systematic
evolution of HBT interferometry for identical pions.
This comprehensive data set establishes
how the extracted HBT parameters vary as
a function of the initial
geometry of the reaction.
Any change in this 
relationship may indicate the onset of new phenomena, e.g.
a change in expansion driven by forming a small volume of 
baryon-rich QGP\cite{Ber89,Pratt90,Ris96}.
 
Each pion two-particle correlation function is measured in three
dimensions corresponding to the three components of 
relative momentum\cite{Ber89,Pratt90}: 
q$_z$ is parallel to the beam, q$_{tout}$ is 
parallel to the sum of the transverse momenta of the pair and q$_{tside}$ 
is in the orthogonal transverse direction.  
The correlation functions were fit with the following Gaussian.
\begin{equation}
C(q)=1+\lambda e^{-((R_{tout}q_{tout})^2+(R_{tside}q_{tside})^2+
(R_{z}q_{z})^2)/2}
\end{equation}
\begin{figure}[htb]
\begin{minipage}[t]{75mm}
\epsfxsize=7.5cm\epsfbox[30 120 410 375]{
%earth$scratch:[ogilvie.qm97]rtside.eps}
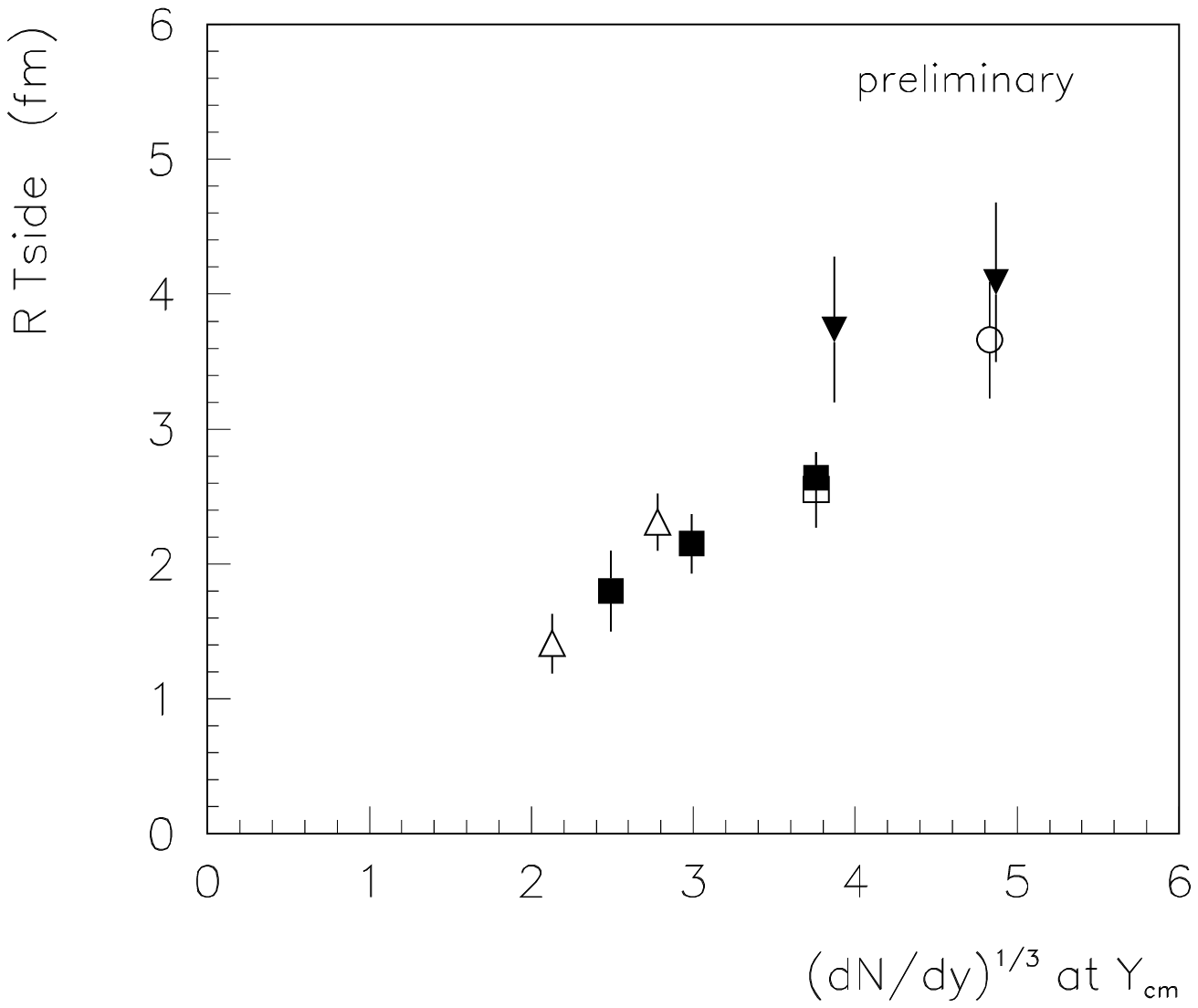}
\end{minipage}
\hspace{\fill}
\begin{minipage}[t]{75mm}
\epsfxsize=7.5cm\epsfbox[30 120 410 375]{
%earth$scratch:[ogilvie.qm97]rdiff.eps}
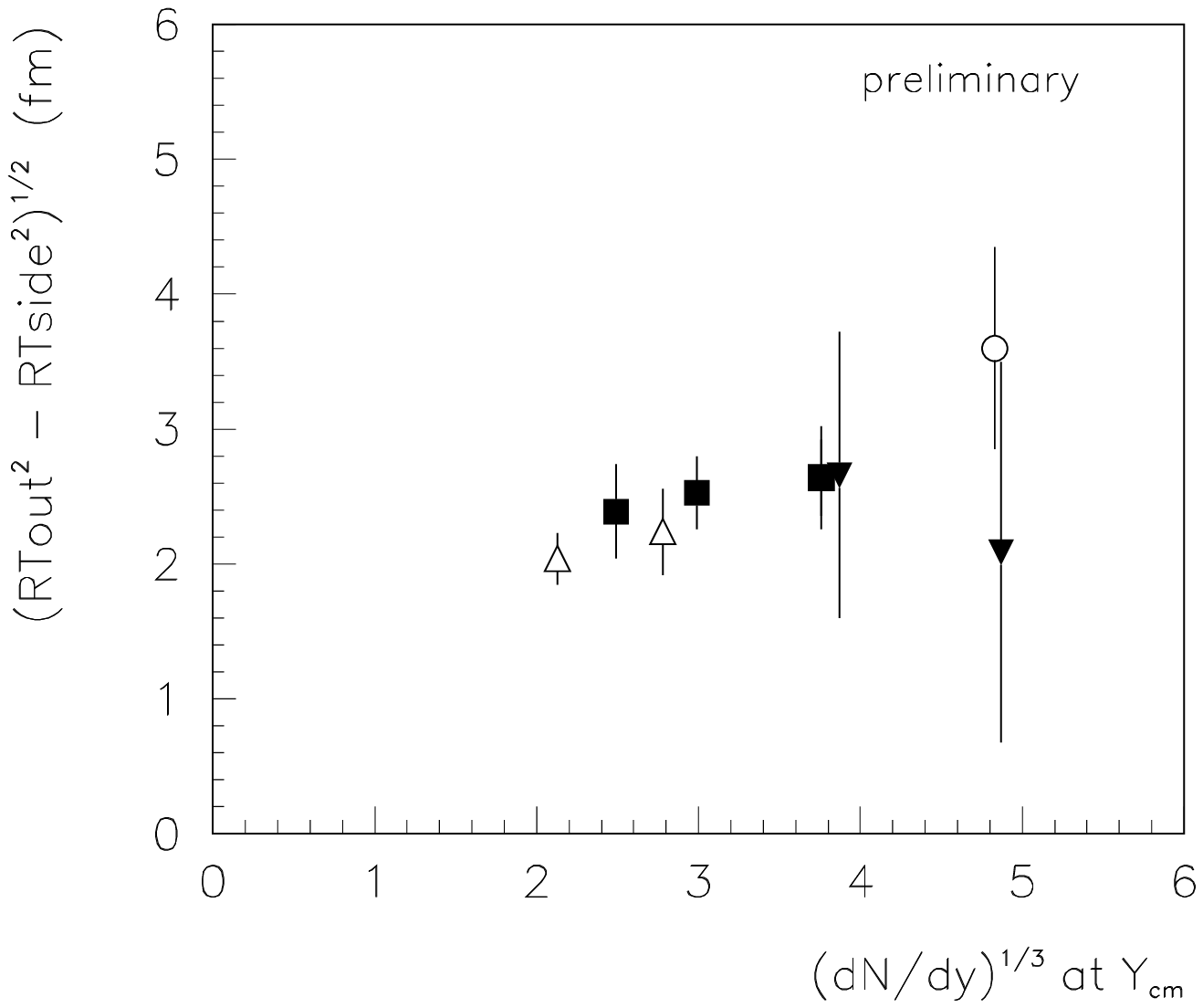}
\end{minipage}
\caption{The extracted parameter R$_{tside}$ (left panel)
and the quadratic difference 
of R$_{tout}$ and R$_{tside}$  (right panel)
versus the cube-root of the total pion dN/dy.
E859 Si+Al, Si+Au results are shown as
open triangles and filled
squares respectively. 
E866 Au+Au results are shown
as open circles and filled inverted triangles.
The errors are statistical only.}
\label{fig:rtside}
\end{figure}                             
The extracted parameter $R_{tside}$ is primarily sensitive to the transverse 
geometry and dynamics of the source.  In the left panel of
Figure~\ref{fig:rtside}, 
R$_{tside}$ shows a steady, linear increase 
with the cube-root of the pion rapidity density.  
There is no anomalous increase that 
would indicate a rapid change in the size of the
emitting system or a change in its expansion dynamics.

The quadratic difference of R$_{tout}$ and R$_{tside}$
(right panel of Figure~\ref{fig:rtside}) is sensitive to 
both the duration of emission of the source and its 
dynamics\cite{Ber89,Pratt90}.  
In contrast to R$_{tside}$, 
the quadratic difference is independent of the initial system size.
This is consistent with the duration of emission being similar
for all the systems studied.

\section{TESTS OF EQUILIBRIUM}
 
It is possible that information from an early, 
small, volume of QGP may 
be lost in the randomizing collisions during the hadronic phase.  Inelastic 
collisions alter the  signatures of particle abundances and all 
collisions alter the kinematic signatures.
A system with a high collision rate will be driven towards
thermal equilibrium and potentially lose information
about a possible QGP.
To assess this possibility, we need to quantify how close the
emitting system is  
to thermal equilibrium.
 
\begin{figure}[htb]
\epsfysize=6.5cm\epsfbox[3 120 450 410]{
%earth$scratch:[ogilvie.qm97]dndy_rat.eps}
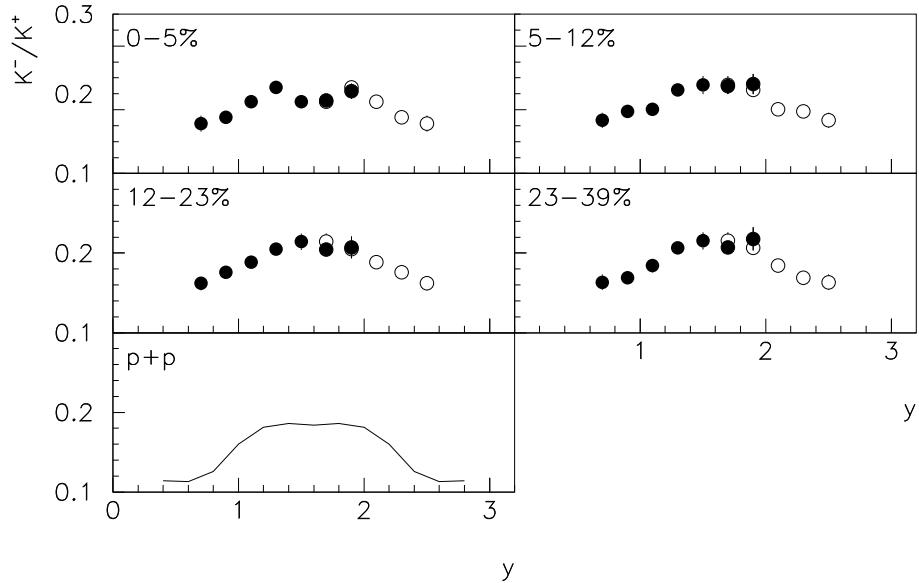}
\caption{The ratio of the 
the K$^-$ and K$^+$ dN/dy distributions
for different centrality classes in Au+Au reactions at 11.6 A~GeV/c.
The errors are statistical only.
The lower panel is for p+p reactions at 12 GeV/c and is adapted from
Fesefeldt et al. \protect\cite{Fes79}.}
\label{fig:raty}
\end{figure}

As observed earlier in this paper, the rapidity
distributions for K$^+$ and K$^-$ are different. 
This is emphasized in 
Figure~\ref{fig:raty} which shows the 
K$^-$/K$^+$ ratio as function of rapidity for different centrality classes
in Au+Au reactions at 11.6 A~GeV/c.  
The rapidity distribution for K$^-$ is narrower
than the K$^+$ distribution for all centralities and
also for p+p reactions at 12 GeV/c.
This is consistent with K$^-$ 
having less phase space than K$^+$ due to a higher 
energy threshold for K$^-$ production.  

\begin{figure}[htb]
\epsfysize=6.5cm\epsfbox[3 120 450 410]{
%earth$scratch:[ogilvie.qm97]tdiff.eps}
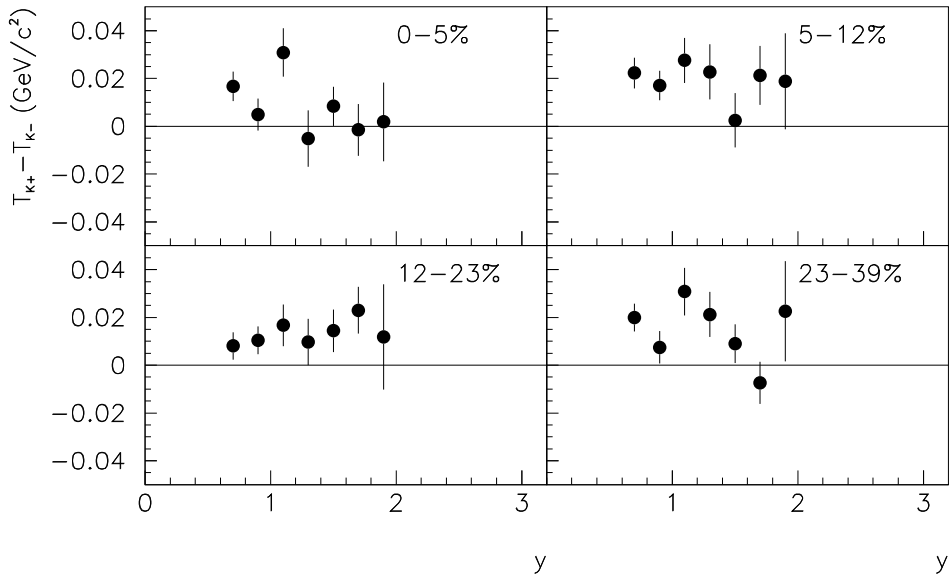}
\caption{The difference between  
the K$^+$ and K$^-$ inverse slope parameters
as a function of rapidity
for different centrality classes in Au+Au reactions at 11.6 A~GeV/c.
The errors are statistical only.}
\label{fig:tdiff}
\end{figure}

Figure~\ref{fig:tdiff} shows the 
difference between the inverse slopes from K$^+$ and K$^-$ 
transverse spectra as a function of rapidity for each centrality class
in Au+Au reactions at 11.6 A~GeV/c.
Systematically the K$^+$ inverse slopes are 10 to 20 MeV
larger than the K$^-$ inverse slopes.    
The difference in inverse slopes is consistent with
less phase space available for K$^-$ production.

The fact that K$^+$ and K$^-$ have different rapidity distributions
and transverse spectra excludes a commonly used class of 
thermal models\cite{PBM}, namely those models
that assume that 
thermal parameters are uniform across the emitting system
and that all particles are in kinetic equilibrium.
For a system at kinetic equilibrium, there is sufficient
scattering so that                        
the spectra of equal mass particles are identical in shape,
even if the system is expanding.
It may however be possible to reconcile these data
with a modified thermal model, either with a 
rapidity-dependent chemical potential or temperature,
or possibly different freezeout conditions for K$^-$ and K$^+$.

The comparison between the K$^+$ and K$^-$ spectra may also
constrain and test hadronic models that include
in-medium properties of kaons.  
The effects of a
reduced 
kaon mass are often modeled by an attractive 
kaon mean-field\cite{Li96,Ehe96}.  
This would pull K$^-$ particles to lower m$_t$ and reduce the inverse 
slope parameter.  
The repulsive mean-field for K$^+$ should increase 
the inverse slope parameter.
Because the mean-fields are predicted to increase with 
density, the difference in kaon spectral shapes 
might be expected to grow for more central 
reactions.  This is not observed.

\section{Summary of Results and Conclusions}

By mapping out the evolution of 
the spectra, yields and correlation
functions of produced particles,
we have probed the effects of multiple hadronic
collisions,
absorption, and energy thresholds in heavy-ion reactions.  
Across the broad systematic data set
there is no evidence for any
onset of new behavior beyond hadronic scattering as the 
beam energy or centrality is changed. 
There are however three outstanding puzzles;
1) why is the ratio of yields K$^-$/K$^+$ independent
of centrality, 2) why do the transverse spectra
of $\overline{\rm{p}}$ and protons have similar shapes,
and 3) why is the quadratic difference of R$_{tout}$ and
R$_{tside}$ constant?

One of the largest puzzles in AGS physics is
that hadronic models\cite{Sorge89,ART,ARC} predict 
that a small region of very dense nuclear matter
($\rho > 8\rho_0$)
exists during an Au+Au collision, but no evidence
for a QGP has been observed.  
There are two 
possibilities: 1) a baryon-rich QGP is not formed in such a region
or is very small, 2) the
information from such a plasma is lost after hadronization by the many hadronic
collisions.  
The second possibility makes it difficult to
set a quantitative limit on QGP formation.
To do so will require a model-dependent study
of inserting an ad-hoc change into the particle distributions
during the model
evolution of a heavy-ion reaction.
How much information is retained from this disturbance  
provides an estimate of the survivability of QGP signatures.
This clearly depends on how frequent the collisions are in the
hadronic phase.
The measured spectral differences between K$^+$ and K$^-$  
can be used to test the assumptions of hadronic collision rates 
in such a study. 

Experimentally, the challenge is to extend the results to
more sensitive probes, such as the yield of
$\overline{\Lambda}$, $\phi$, and also to extend our studies
to more exclusive variables, e.g. the 
azimuthal distribution of particle production with respect to the reaction
plane.
It will also be important to increase
the range of observables included in the excitation function
to proton stopping via the proton rapidity distributions,
as well as to the rapidity and m$_{T}$ dependence of HBT source
parameters.

This work was supported by the U.S. Department of Energy under contracts
with BNL, Columbia University, LLNL, MIT, UC
Riverside, by NASA, under
contract with the University of California, by Ministry of Education
and KOSEF in Korea, and by the Ministry of Education,
Science, Sports, and Culture of Japan.


\begin{thebibliography}{9}
\bibitem{caveats} Fermi motion in Au will also increase the 
K/$\pi$ double ratio but Fermi motion cannot account for the
increased K/$\pi$ ratio observed as a function of centrality.
\bibitem{feed} The $\overline{\rm{p}}$ have not been corrected
for feed-down from $\overline{\Lambda}$ decay.
For the Henry Higgins and Forward spectrometers the acceptance
for daughter $\overline{\rm{p}}$  from $\overline{\Lambda}$ decay
is approximately 100\% and 65\%, respectively.
\end{thebibliography}
\end{document}